\documentclass[preprint,10pt]{elsarticle}                     
\usepackage{color}   
\usepackage[bookmarks=true,pdfencoding=unicode]{hyperref}
\hypersetup{
    backref=true,
    citecolor=blue,
    colorlinks=true, 
    linktoc=all,     
    linkcolor=blue,  
}

\usepackage{bookmark}
\usepackage[latin1]{inputenc}
\usepackage[T1]{fontenc}
\usepackage{bezier}
\usepackage{dsfont}
\usepackage{mathptmx}      
\usepackage{amsfonts}
\usepackage{amsmath}
\usepackage{amssymb}
\usepackage{amsbsy}
\usepackage{amscd}
\usepackage{amsopn}
\usepackage{amssymb}
\usepackage{amstext}

\usepackage{amsxtra}
\usepackage{array}
\usepackage{bm}
\usepackage{graphicx,psfrag}
\usepackage{latexsym}
\usepackage{multirow}
\usepackage{subfig}
\usepackage{etoolbox}
\usepackage{algorithm}
\usepackage{algorithmicx}
\usepackage{algpseudocode}
\algtext*{EndIf}
\usepackage{mdframed}
\usepackage{lipsum} 

\DeclareMathAlphabet{\mathcal}{OMS}{cmsy}{m}{n}

\newtoggle{LatexWithProofs}
\togglefalse{LatexWithProofs}

\newtoggle{LatexFull}
\togglefalse{LatexFull}

\newtheorem{pLemma}{Lemma}
\newtheorem{pProp}{Proposition}
\newtheorem{pArgument}{Argument}
\newtheorem{pExample}{Example}
\newtheorem{pCorol}{Corollary}
\newtheorem{pLongDef}{Definition}
\newtheorem{pTheorem}{Theorem}

\newcommand{\pRef}[1]{(\ref{#1})}

\newcommand{\peLongDef}[1]{$\blacktriangle$ \end{pLongDef}}
\newcommand{\peLongDefX}[1]{\end{pLongDef}}

\newcommand{\peArgument}[1]{ \hyperlink{#1}{$\blacktriangle$} \end{pArgument}}

\newcommand{\peExample}[1]{ \hyperlink{#1}{$\blacktriangle$} \end{pExample}}
\newcommand{\peExampleX}[1]{\end{pExample}}

\newcommand{\pePlainExample}[1]{$\blacktriangle$ \end{pExample}}
\newcommand{\pePlainExampleX}[1]{\end{pExample}}

\newcommand{\peCorol}[1]{\hyperlink{#1}{$\blacktriangle$} \end{pCorol}}
\newcommand{\peCorolX}[1]{\end{pCorol}}

\newcommand{\peLemma}[1]{ \hyperlink{#1}{$\blacktriangle$} \end{pLemma}}
\newcommand{\peLemmaX}[1]{\end{pLemma}}

\newcommand{\peProp}[1]{ \hyperlink{#1}{$\blacktriangle$} \end{pProp}}
\newcommand{\pePropX}[1]{\end{pProp}}

\newcommand{\peTheorem}[1]{ \hyperlink{#1}{$\blacktriangle$} \end{pTheorem}}

\newcommand{\pLink}[1]{\eqno \hyperlink{#1}{\blacktriangle}}

\iftoggle{LatexFull}
{
\newcommand{\pFullLink}[1]{\eqno \hyperlink{#1}{\blacktriangle}}
\newcommand{\peFullProp}[1]{ \hyperlink{#1}{$\blacktriangle$} \end{pProp}}
\newcommand{\peFullCorol}[1]{ \hyperlink{#1}{$\blacktriangle$} \end{pCorol}}
\newcommand{\peFullLemma}[1]{ \hyperlink{#1}{$\blacktriangle$} \end{pLemma}}
\newcommand{\peFullExample}[1]{ \hyperlink{#1}{$\blacktriangle$} \end{pExample}}
}
{
\newcommand{\pFullLink}[1]{\eqno \blacktriangle}
\newcommand{\peFullProp}[1]{$\blacktriangle$ \end{pProp}}
\newcommand{\peFullCorol}[1]{$\blacktriangle$ \end{pCorol}}
\newcommand{\peFullLemma}[1]{$\blacktriangle$ \end{pLemma}}
\newcommand{\peFullExample}[1]{$\blacktriangle$ \end{pExample}}
}

\iftoggle{LatexWithProofs}
{
\definecolor{darkpink}{rgb}{0.91, 0.33, 0.5}
\definecolor{darksalmon}{rgb}{0.91, 0.59, 0.48}
\definecolor{desertsand}{rgb}{0.93, 0.79, 0.69}
\definecolor{celadon}{rgb}{0.67, 0.88, 0.69}
\definecolor{darkcyan}{rgb}{0.0, 0.55, 0.55}

\newcommand{\pbClaim}[1]{{\color{red} ?}}
\newcommand{\peClaim}[1]{{\color{red} ?`}}
\newcommand{\pbArgument}[1]{\begin{pArgument} \label{#1}  {\color{darksalmon} #1}}

\newcommand{\pbExample}[1]{\begin{pExample} \label{#1}  {\color{darksalmon} #1}}

\newcommand{\pbExampleB}[1]{
\begin{pExample} \label{#1}
{\color{darksalmon} #1}
\hypertarget{{T#1}}{}
\bookmark[
rellevel=1,
keeplevel,
dest=T#1
]{Example \ref{#1}}
}

\newcommand{\pbExampleBT}[2]{
\begin{pExample}[#2] \label{#1}
{\color{darksalmon} #1}
\hypertarget{{T#1}}{}
\bookmark[
rellevel=1,
keeplevel,
dest=T#1
]{Example \ref{#1}: {#2}}
}

\newcommand{\pbCorol}[1]{\begin{pCorol} \label{#1}  {\color{darksalmon} #1}}

\newcommand{\pbCorolB}[1]{
\begin{pCorol} \label{#1}  {\color{darksalmon} #1}
\hypertarget{{T#1}}{}
\bookmark[
rellevel=1,
keeplevel,
dest=T#1
]{Corollary \ref{#1}}
}

\newcommand{\pbCorolBT}[2]{
\begin{pCorol}[#2]\label{#1}  {\color{darksalmon} #1}
\hypertarget{{T#1}}{}
\bookmark[
rellevel=1,
keeplevel,
dest=T#1
]{Cor. \ref{#1}: {#2}}
}

\newcommand{\pbLemma}[1]{\begin{pLemma} \label{#1}  {\color{darksalmon} #1}}

\newcommand{\pbLemmaB}[1]{
\begin{pLemma} \label{#1}  {\color{darksalmon} #1}
\hypertarget{{T#1}}{}
\bookmark[
rellevel=1,
keeplevel,
dest=T#1
]{Lemma \ref{#1}}
}

\newcommand{\pbLemmaBT}[2]{
\begin{pLemma}[#2] \label{#1}  {\color{darksalmon} #1}
\hypertarget{{T#1}}{}
\bookmark[
rellevel=1,
keeplevel,
dest=T#1
]{Lem. \ref{#1}: {#2}}
}

\newcommand{\pbProp}[1]{\begin{pProp} \label{#1}  {\color{darksalmon} #1}}

\newcommand{\pbPropB}[1]{
\begin{pProp} \label{#1}  {\color{darksalmon} #1}
\hypertarget{{T#1}}{}
\bookmark[
rellevel=1,
keeplevel,
dest=T#1
]{Prop. \ref{#1}}
}

\newcommand{\pbPropBT}[2]{
\begin{pProp}[#2] \label{#1}  {\color{darksalmon} #1}
\hypertarget{{T#1}}{}
\bookmark[
rellevel=1,
keeplevel,
dest=T#1
]{Prop. \ref{#1}: {#2}}
}

\newcommand{\pbTheorem}[1]{\begin{pTheorem} \label{#1}  {\color{darksalmon} #1}}

\newcommand{\pbTheoremB}[1]{
\begin{pTheorem} \label{#1}  {\color{darksalmon} #1}
\hypertarget{{T#1}}{}
\bookmark[
rellevel=1,
keeplevel,
dest=T#1
]{Theorem \ref{#1}}
}

\newcommand{\pbTheoremBT}[2]{
\begin{pTheorem}[#2] \label{#1}  {\color{darksalmon} #1}
\hypertarget{{T#1}}{}
\bookmark[
rellevel=1,
keeplevel,
dest=T#1
]{Theor. \ref{#1}: {#2}}
}

\newcommand{\pbLongDef}[1]{\begin{pLongDef}
\label{#1} \hypertarget{{#1}}{} {\color{darksalmon} #1}
\bookmark[
rellevel=1,
keeplevel,
dest=#1
]{Definition \ref{#1}}
}

\newcommand{\pbLongDefB}[2]{\begin{pLongDef}
\label{#1} \hypertarget{{#1}}{} {\color{darksalmon} #1}
\bookmark[
rellevel=1,
keeplevel,
dest=#1
]{\ref{#1}: #2}
}

\newcommand{\pbLongDefBT}[2]{\begin{pLongDef}[#2]
\label{#1} \hypertarget{{#1}}{} {\color{darksalmon} #1}
\bookmark[
rellevel=1,
keeplevel,
dest=#1
]{\ref{#1}: #2}
}

\newcommand{\pbChain}[1]{\[ }
\newcommand{\peChain}[1]{\ {\color{red} \checkmark  \wrm{#1}} \] }

\newcommand{\pePClaim}[1]{ \checkmark}

\newcommand{\pbTClaim}[1]{\begin{equation} \label{#1}  }
\newcommand{\peTClaim}[1]{ \ {\color{red} \checkmark  \wrm{#1}} \end{equation} }
\newcommand{\peEClaim}[1]{ \ {\color{red} \checkmark  \wrm{#1 \bullet} }\end{equation} }
\newcommand{\pClaimLabel}[1]{ \ {\label{#1} \color{red} \checkmark  \wrm{#1}}}

\newcommand{\pbDef}[1] {\hypertarget{{#1}}{} \begin{equation} \label{#1} }
\newcommand{\peDef}[1]{ {\color{blue} \bigstar \wrm{#1}} \end{equation}}

\newcommand{\pbHypot}[1]{ \begin{equation} \label{#1}  }
\newcommand{\peHypot}[1]{  \ \ {{\color{blue} \bigstar \wrm{#1}}}  \end{equation} }

\newcommand{\pbProof}[2]{\newpage {\color{darksalmon} Proof of {#1} {#2}} \ref{#2} \hypertarget{{#2}}{}}
\newcommand{\pbProofB}[2]{\newpage {\color{darksalmon} Proof of {#1} {#2}} \ref{#2} \hypertarget{{#2}}{}
\bookmark[
rellevel=1,
keeplevel,
dest=#2
]
{{#1} \ref{#2}}
}

\newcommand{\peProof}[2]{\qed{} {\color{darksalmon} End of proof of #1 #2} \ref{#2}}

\newcommand{\peVerify}[1]{\qed{} {\color{darksalmon} End of verification of Example #1} \ref{#1}}

}
{
\newcommand{\pbChain}[1]{\[}
\newcommand{\peChain}[1]{\]}

\newcommand{\pbClaim}[1]{}
\newcommand{\peClaim}[1]{}
\newcommand{\pClaimLabel}[1]{\label{#1}}
\newcommand{\peTClaim}[1]{ \end{equation}}
\newcommand{\peEClaim}[1]{ \end{equation}}
\newcommand{\pbTClaim}[1]{\begin{equation} \label{#1}}

\newcommand{\peDef}[1]{  \end{equation}}

\newcommand{\lpPlainClaim}[1]{}
\newcommand{\lpEndPlainClaim}[1]{}
\newcommand{\pbDef}[1]{ \hypertarget{{#1}}{} \begin{equation} \label{#1}}

\newcommand{\pbHypot}[1]{\begin{equation} \label{#1}  }
\newcommand{\peHypot}[1]{ \end{equation}}

\newcommand{\pbCorol}[1]{\begin{pCorol} \label{#1}}

\newcommand{\pbCorolB}[1]{\begin{pCorol} \label{#1}
\hypertarget{{T#1}}{}
\bookmark[
rellevel=1,
keeplevel,
dest=T#1
]
{Corollary \ref{#1}}
}

\newcommand{\pbCorolBT}[2]{
\begin{pCorol}[#2] \label{#1}
\hypertarget{{T#1}}{}
\bookmark[
rellevel=1,
keeplevel,
dest=T#1
]
{Cor. \ref{#1}: {#2}}
}

\newcommand{\pbArgument}[1]{\begin{pArgument} \label{#1}}
\newcommand{\pbExample}[1]{\begin{pExample} \label{#1}}

\newcommand{\pbExampleB}[1]{\begin{pExample} \label{#1}
\hypertarget{{T#1}}{}
\bookmark[
rellevel=1,
keeplevel,
dest=T#1
]{Example \ref{#1}}
}

\newcommand{\pbExampleBT}[2]{
\begin{pExample}[#2] \label{#1}
\hypertarget{{T#1}}{}
\bookmark[
rellevel=1,
keeplevel,
dest=T#1
]{Example \ref{#1}: {#2}}
}

\newcommand{\pbLemma}[1]{\begin{pLemma} \label{#1}}
\newcommand{\pbLemmaBT}[2]{\begin{pLemma}[#2] \label{#1}
\hypertarget{{T#1}}{}
\bookmark[
rellevel=1,
keeplevel,
dest=T#1
]
{Lem. \ref{#1}: {#2}}
}

\newcommand{\pbLemmaB}[1]{\begin{pLemma} \label{#1}
\hypertarget{{T#1}}{}
\bookmark[
rellevel=1,
keeplevel,
dest=T#1
]
{Lemma \ref{#1}}
}

\newcommand{\pbProp}[1]{\begin{pProp} \label{#1}}
\newcommand{\pbPropBT}[2]{\begin{pProp}[#2] \label{#1}
\hypertarget{{T#1}}{}
\bookmark[
rellevel=1,
keeplevel,
dest=T#1
]
{Prop. \ref{#1}: {#2}}
}

\newcommand{\pbPropB}[1]{\begin{pProp} \label{#1}
\hypertarget{{T#1}}{}
\bookmark[
rellevel=1,
keeplevel,
dest=T#1
]
{Prop. \ref{#1}}
}

\newcommand{\pbTheorem}[1]{\begin{pTheorem} \label{#1}}
\newcommand{\pbLongDef}[1]{\begin{pLongDef} \label{#1}}

\newcommand{\pbLongDefB}[2]{\begin{pLongDef}
\label{#1} \hypertarget{{#1}}{}
\bookmark[
rellevel=1,
keeplevel,
dest=#1
]{\ref{#1}: #2}
}

\newcommand{\pbLongDefBT}[2]{\begin{pLongDef}[#2]
\label{#1} \hypertarget{{#1}}{}
\bookmark[
rellevel=1,
keeplevel,
dest=#1
]{\ref{#1}: #2}
}

\newcommand{\pbProof}[2]{ {\bf Proof of #1 \ref{#2}}  \hypertarget{{#2}}}

\newcommand{\pbProofB}[2]{ {\bf Proof of {#1} \ref{#2}} \hypertarget{{#2}}{}
\bookmark[
rellevel=1,
keeplevel,
dest=#2
]
{{#1} \ref{#2}}
}

\newcommand{\peProof}[2]{\qed{}}

\newcommand{\peVerify}[1]{\qed{}}
}

\newcommand{\wabs}[1]{\left|#1\right|}

\newcommand{\wcal}[1]{\mathcal{#1}}

\newcommand{\wceil}[1]{\lceil #1 \rceil}

\newcommand{\wfc}[2]{{#1}\!\left(#2\right)}

\newcommand{\wfl}[1]{\wfc{\wrm{fl}}{#1}}

\newcommand{\wflr}[2]{\wfc{\wrm{#1}}{#2}}

\newcommand{\wvfc}[2]{{\wvec{#1}}\!\left(#2\right)}

\newcommand{\wfloor}[1]{\lfloor {{#1}} \rfloor }

\newcommand{\wi}[1]{\wrm{i}}

\algdef{SE}[DOWHILE]{Do}{doWhile}{\algorithmicdo}[1]{\algorithmicwhile\ #1}%

\newdimen\CdotAxis
\newcommand*{\CdotAux}[3]{%
  {%
    \settoheight\CdotAxis{$#2\vcenter{}$}%
    \sbox0{%
      \raisebox\CdotAxis{%
        \scalebox{#1}{%
          \raisebox{-\CdotAxis}{%
            $\mathsurround=0pt #2#3$%
          }%
        }%
      }%
    }%
    \dp0=0pt %
    \sbox2{$#2\bullet$}%
    \ifdim\ht2<\ht0 %
      \ht0=\ht2 %
    \fi
    \sbox2{$\mathsurround=0pt #2#3$}%
    \hbox to \wd2{\hss\usebox{0}\hss}%
  }%
}

\newcommand{\wlr}[1]{\left( #1 \right)}

\newcommand{\wnorm}[1]{\left\| #1 \right\|}
\newcommand{\wvnorm}[2]{{\left\| \wvec{#1}_{#2} \right\|}}

\newcommand{\wrone}{\mathds R}
\newcommand{\wrn}[1]{{\mathds R}^{#1}}
\newcommand{\wrm}[1]{\mathrm{#1}}

\newcommand{\wset}[1]{{\left\{ #1 \right\}}}

\newcommand{\wsign}[1]{\wfc{\wrm{sign}}{#1}}

\newcommand{\wvec}[1]{\mathbf{#1}}

\newcommand{\wfpa}{\alpha}

\newcommand{\wfpes}[1]{\wcal{E}_{\wcal{A},e}}

\newcolumntype{M}[1]{>{\centering\arraybackslash}m{#1}}
\newcolumntype{N}{@{}m{0pt}@{}}

\begin{document}
\begin{frontmatter}
\title{Fast and accurate normalization of vectors and quaternions}
\author{Walter F. Mascarenhas}
\ead{walter.mascarenhas@gmail.com}
\address{ Instituto de Matem\'{a}tica e Estat\'{i}stica, Universidade de S\~{a}o Paulo, Brazil
\fnref{fna}}
\fntext[fna]{Depto. de Computa\,{c}\~{a}o, IME USP,
      Cidade Universit\'{a}ria, Rua do Mat\~{a}o 1010, S\~{a}o Paulo SP, Brazil. CEP 05508-090,
         Tel.: +55-11-3091 5411, Fax: +55-11-3091 6134}

\begin{abstract}
We present fast and accurate ways to normalize
two and three dimensional vectors and quaternions and compute their length.
Our approach is an adaptation of ideas used in the
linear algebra library LAPACK, and we believe that
the computational geometry and computer aided design communities
are not aware of the possibility of speeding up these
fundamental operations in the robust way proposed here.
\end{abstract}


\end{frontmatter}

\section{Introduction}
Obtaining the length of vectors and normalizing them
are basic operations implemented by most computational geometry and
computer aided design software libraries. In this case,
for $n = 2$ or $3$, given a vector $\wvec{x} \in \wrn{n} \setminus \wset{0}$  we compute
\pbDef{normA}
\wvnorm{x}{} := \sqrt{x_1^2 + \dots + x_n^2}
\hspace{1.0cm} \wrm{and} \hspace{1.0cm}
\bar{\wvec{x}} := \wlr{\frac{x_1}{r}, \dots, \frac{x_n}{r}},
\peDef{normA}
so that $\bar{\wvec{x}}$ has length one
and can be used in geometric primitives, like the
computation of the distance of a point to a line
or a plane. In the ObjectARX$^\circledR$ library \cite{ObjectARX} for example, vectors
have a normalize method and
most libraries for computational geometry and CAD have
similar functions.

Sums of squares are also required for dealing with quaternions, which
are used to represent rotations in three dimensions. In fact,
when we represent rotations by quaternions we either assume that they are normalized
or perform a division by the square
of their norm. Therefore, when dealing with a quaternion
$\wvec{q} := \wlr{x_1,x_2,x_3,x_4}$
%
\pbDef{normQ}
x_1^2 + x_2^2 + x_3^2 + x_4^2
\hspace{1cm} \wrm{or} \hspace{1cm}
\wfc{r}{\wvec{q}} := \wvnorm{q}{} := \sqrt{x_1^2 + x_2^2 + x_3^2 + x_4^2},
\peDef{normQ}
and even when we only apply rotations with normalized quaternions
we must evaluate the expressions in Equation \pRef{normQ}
in the preliminary step in which we compute them.

This note presents efficient and accurate algorithms
to evaluate the expressions in Equations \pRef{normA}--\pRef{normQ}.
In theory, evaluating these expressions is easy,
we need only to pay attention to $\wvec{x} = 0$.
The robust implementation of these expressions
in finite precision arithmetic is also well known among
experts in computational geometry. For example, we wrote to
a respected scholar in this field and received
a precise answer, explaining that actually
normalization of vectors is a bit tricky, because when $x_1$ is too small
the computed value $\wfl{x_1^2}$ may be zero even
when $x_1$ is different from zero, and similarly
$\wfl{x_1^2}$ may overflow when $x_1$ is large.
The authors of professional software libraries
are also well aware of these problems.
The documentation of the
ObjectARX$^\circledR$ library
for instance states clearly that the normalization may fail
when the vector $\wvec{x}$ is too small.

\begin{algorithm}
\caption{The robust Quotient Algorithm for normalizing three dimensional vectors}
\label{algoQuotThree}
\begin{algorithmic}[1]
\State $m \gets \max\wset{ \wabs{x_1}, \wabs{x_2}, \wabs{x_3}}$
\If{ $m = 0$ }
\State $r \gets 0$.
\Else
\State $x_1 \gets x_1/m$, $\ \ \  x_2 \gets x_2 /m$, $\ \ \ x_3 \gets x_3 / m$
\State $r \gets m \sqrt{x_1 * x_1 + x_2 * x_2 + x_3 * x_3}$
\State $x_1 \gets x_1/r$, $\ \ \  x_2 \gets  x_2 /r$, $\ \ \ x_3 \gets x_3 / r$
\EndIf
\end{algorithmic}
\end{algorithm}

The scholar mentioned that Algorithm \ref{algoQuotThree}
above would handle overflow, and we could also use Algorithm \ref{algoQuotThreeB}
below. These algorithms use divisions, and
we call them ``quotient algorithms.''
When the input is finite, they only fail due to overflow, in situations
in which their input warrants this result.
Moreover, Algorithm 2 returns NaNs (not a numbers) when the input contains NaNs.

\begin{algorithm}[h]
\caption{The fast and robust Quotient Algorithm for normalizing 3D vectors}
\label{algoQuotThreeB}
\begin{algorithmic}[1]
\State \Comment Input: $x_1,x_2,x_3$, Output $[r  , \, x_1/r, \, x_2/r, \,x_3/r]$ for $r = \sqrt{x_1^2 + x_2^2 + x_3^2}$
\If{$\wabs{x_1} > \wabs{x_2}$ }
    \If{ $\wabs{x_3} > \wabs{x_1}$ }
        \State $q_1 \gets x_1/x_3$, $\ \ \ \ q_2 \gets x_2/x_3$
        \State $h \gets \sqrt{1 + q_1 * q_1 + q_2 * q_2}$
        \State $r \gets \wabs{x_3} h$
        \State $x_3 \gets \wsign{x_3} / h$
        \State \Return $[r, q_1 * x_3, \,  q_2 * x_3, \, x_3]$
    \Else
        \State Proceed as in lines 4--8, with $x_1$ playing the role of $x_3$ and vice versa
    \EndIf
\Else
    \If{ $\wabs{x_3} \geq \wabs{x_2}$ }
        \If{ $x_3 = 0$} \Return [0,0,0,0] \EndIf
        \State Goto line 4
    \Else
        \State Proceed as in lines 4--8, with $x_2$ playing the role of $x_3$ and vice versa
    \EndIf
\EndIf
\end{algorithmic}
\end{algorithm}

We have been using Algorithm \ref{algoQuotThreeB} for many years, but
found recently that by adapting the
method proposed by Bindel, Demmel, Kahan and Marques in \cite{Bindel}
to compute plane rotations we obtain a better way to evaluate the expressions
in Equations \pRef{normA}--\pRef{normQ}.
Their method uses scaling instead of divisions and
leads to the ``scaling algorithms'' presented in this article.
The superiority of scaling algorithms to quotient algorithms
is illustrated by Table \ref{tableTimes} in \ref{secExperiments}.
The degree by which a scaling algorithm is superior to the corresponding
quotient algorithm in the entries of Table \ref{tableTimes}
varies, but cases in which the scaling algorithm is twice as fast as
the quotient algorithm are not uncommon. Moreover, there is no
instance in which the scaling algorithm is slower, and  
we prove here that the rounding errors incurred by scaling 
algorithms are small.

One may ask whether it wouldn't be simpler to code
Equations \pRef{normA}--\pRef{normQ} in the obvious way.
This question is relevant, specially when one knows
that underflow and overflow will not happen. Table \ref{tablePlain} in \ref{secExperiments}
may help one
to decide whether the robustness brought by scaling is worth its cost.
This table compares the algorithms described here with the
naive algorithm. On the one hand, it shows that
scaling can be expensive. For instance,
we frequently normalize quaternions which already have
a norm close to one, and the last two rows
of Table \ref{tablePlain} show that
scaling is not worth in this case.
On the other hand, Table \ref{tablePlain}
also shows that, depending on the compiler and the precision, there
are cases in which the overhead due to scaling is minimal.

In the rest of this note we present scaling algorithms
to normalize vectors and quaternions and compute their length,
and describe their numerical properties. Section \ref{secVectors} describes
scaling algorithms for normalizing vectors
and computing their norm, and presents formal results
about the numerical properties of these algorithms.
Section \ref{secQuaternions} does the same for quaternions.
\ref{secExperiments} contains the experimental results
and in \ref{secProofs} we prove that our estimates
for the rounding errors are correct.

\section{Accurate and efficient scaling}
\label{secScaling}
Quotient and scaling algorithms have the same motivation: to
avoid underflow and overflow in the evaluation of
\pbDef{squares}
x_1^2 + x_2^2,
\hspace{0.9cm}
x_1^2 + x_2^2 + x_3^2
\hspace{0.9cm} \wrm{and} \hspace{0.9cm}
x_1^2 + x_2^2 + x_3^2 + x_4^2.
\peDef{squares}
They both evaluate $m := \max \wabs{x_k}$, but use $m$
differently. Quotient algorithms divide $\wvec{x}$ by $m$, and
are inefficient because
divisions are much more expensive than sums and multiplications.
Scaling algorithms perform the minimal
work required to avoid underflow and overflow
in Equation \pRef{squares}. They use
floating point numbers $\tau_{\min}$ and $\tau_{\max}$ such
that if $\tau_{\min} \leq m \leq \tau_{\max}$ then the sums of squares in
\pRef{squares} can be computed with a small relative error.
For instance, the results about dot products in
\cite{RealNumbers} show that, in a binary IEEE machine,
with the parameters in Table \ref{tableParam} below, if

\begin{table}
\caption{The parameters describing a floating point arithmetic}
\centering
\begin{tabular}{M{1.5cm}|M{8.56cm}N}
\hline
Parameter    &  Meaning & \\
\hline
$u$      & unit roundoff & \\
$\alpha$ & the smallest positive floating point number & \\
$\nu$    & the smallest positive normal floating point number & \\
$\omega$ & the largest finite floating point number
\end{tabular}
\label{tableParam}
\end{table}

\pbDef{tau}
\tau_{\min} := 2^{\wceil{\wlr{\wfc{\log_2}{\wfpa/ u^2}}/2}} \hspace{1cm} \wrm{and} \hspace{1cm}
\tau_{\max} := 2^{\wfloor{\wlr{\wfc{\log_2}{\omega} - 3}/2}}
\peDef{tau}
and $\tau_{\min} \leq m \leq \tau_{\max}$ then
we have $\wabs{ \wfl{\wvnorm{x}{}} - \wvnorm{x}{}}  \leq 2.5 \, u \, \wvnorm{x}{}$.
This means that the computed value of $\wvnorm{x}{}$,
which we denote by $\wfl{\wvnorm{x}{}}$,  has a small relative error,
because in single precision $u \approx 6.0 \times 10^{-8}$ and in double precision $u \approx 1.1 \times 10^{-16}$.

\begin{table}
\caption{The parameters for the binary IEEE arithmetics}
\centering
\begin{tabular}{M{1.15cm}|M{0.56cm}M{0.55cm}M{0.55cm}M{0.65cm}|M{1.2cm}M{1.3cm}M{1.3cm}M{1.25cm}N}
             &  \multicolumn{4}{c}{Algorithms} & \multicolumn{4}{c}{Hardware} & \\[0.02cm]
\hline
Precision    &  $\tau_{\min}$    &  $\sigma_{\min}$ & $\tau_{\max}$ & $\sigma_{\max}$ & $u$ & $\alpha$ & $\nu$ & $\omega$ &\\[0.12cm]
\hline
Single       &   $2^{-49}$  & $2^{100}$ &  $2^{62}$  & $2^{-66} $ & $ \approx 10^{-7}$   & $\approx 10^{-45}$  & $\approx  10^{-38}$ & $ \approx  10^{38}$  & \\[0.12cm]
Double       &   $2^{-482}$ & $2^{592}$ &  $2^{510}$ & $2^{-514}$ & $ \approx 10^{-16}$  & $\approx 10^{-324}$ & $\approx  10^{-308}$ & $ \approx  10^{308}$  &
\end{tabular}
\label{tableIEEE}
\end{table}

The key idea is to scale $\wvec{x}$ so that $m$ falls in the
favorable range $[\tau_{\min},\tau_{\max}]$.
To avoid rounding errors, we scale by multiplying by a power of the base of the floating arithmetic,
as when we use the parameters in Table \ref{tableIEEE}.
These parameters satisfy Equations \pRef{condA}--\pRef{condF} and
this ensures that the algorithms for normalizing vectors and quaternions
in the next sections are accurate.
\begin{eqnarray}
\pClaimLabel{condA}
\wfl{\nu^2} = 0, \hspace{1cm} u^2 \geq 16 \wfpa, \hspace{1cm} u \leq 10^{-6}, \\
\pClaimLabel{condB}
\sigma_{\min} \ \wrm{and} \ \sigma_{\max} \ \wrm{are \ powers \ of \ } 2, \\
\pClaimLabel{condC}
 u^2 \tau_{\min}^2 \geq \wfpa \hspace{1cm} \wrm{and} \hspace{1cm} 8 \tau_{\max}^2 \leq \omega, \\
\pClaimLabel{condD}
\omega \tau_{\min}  \geq 1  \ \wrm{ \ and \ } 3  \nu \tau_{\max} \leq 1, \\
\pClaimLabel{condE}
\wrm{if } \  x \in (0,\tau_{\min}] \  \wrm{is \ a \ floating \ point \ number \ then \ } \sigma_{\min} x \in [\tau_{\min},\tau_{\max}], \\
\pClaimLabel{condF}
\wrm{if } \  x \in [\tau_{\max},\omega] \  \wrm{ \ then \ } \sigma_{\max} x \in [\tau_{\min}, \tau_{\max}].
\end{eqnarray}
Algorithms \ref{algoScale2} and \ref{algoScale3} below
scale vectors, so that their largest entry falls in the interval $[\tau_{\min},\tau_{\max}]$.
These algorithms return the inverse of the scaling factor, so that the scaling can be undone afterwards.
Finally, they handle the case in which $\wvec{x} = 0$ properly and
propagate NaNs, in the sense that $x_k$ is NaN for some $k$ if and only if
some of the output $x_k$ is NaN.

\begin{algorithm}[H]
\caption{Scaling two dimensional vectors when $\tau_{\min}$, $\sigma_{\min}$, $\tau_{\max}$ and $\sigma_{\max}$ exist.}
\label{algoScale2}
\begin{algorithmic}[1]
\Procedure{Scale2D}{$x_1$, $x_2$}
\State \Comment Output: $[\sigma^{-1}, \,  \sigma x_1, \, \sigma x_2]$ so that
$\sigma \max \wset{\wabs{x_1},\wabs{x_2}} \in [\tau_{\min},\tau_{\max}]$.
\State $m \gets \wabs{x_1}$
\If{ $m \geq \wabs{x_2}$ }
    \If{ $m = 0$ } \Return $[0,0,0]$ \EndIf
\Else
    \State $m \gets \wabs{x_2}$
\EndIf
\State
\If{ $m \geq \tau_{\min}$ }
    \If{ $m \leq \tau_{\max}$ } \Return $[1, \, x_1, \, x_2]$ \EndIf
    \State  \Return $[\sigma_{\max}^{-1},\,  \sigma_{\max} * x_1, \, \sigma_{\max} * x_2]$
\Else
    \State \Return $[\sigma_{\min}^{-1}, \, \sigma_{\min} * x_1, \, \sigma_{\min} * x_2]$
\EndIf
\EndProcedure
\end{algorithmic}
\end{algorithm}

\begin{algorithm}
\caption{Scaling three dimensional vectors when $\tau_{\min}$, $\sigma_{\min}$, $\tau_{\max}$ and $\sigma_{\max}$ exist.}
\label{algoScale3}
\begin{algorithmic}[1]
\Procedure{Scale3D}{$x_1$, $x_2$, $x_3$}
\State \Comment Output: $[\sigma^{-1}, \,  \sigma x_1, \, \sigma x_2, \, \sigma x_3]$
 so that $\sigma \max \wset{\wabs{x_1},\wabs{x_2},\wabs{x_3}} \in [\tau_{\min},\tau_{\max}]$.
\State $m \gets \wabs{x_1}$
\If{ $m < \wabs{x_2}$ }
    \State $m \gets \wabs{x_2}$
    \If{ $m < \wabs{x_3}$  }
        \State $m \gets \wabs{x_3}$
    \EndIf
\Else
    \If{ $m \geq \wabs{x_3}$ }
        \If{ $m = 0$ } \Return $[0,0,0,0]$ \EndIf
    \Else
        \State $m \gets \wabs{x_3}$
    \EndIf
\EndIf
\State
\If{ $m \geq \tau_{\min}$ }
    \If{ $m \leq \tau_{\max}$ } \Return $[1, \, x_1, \, x_2, \, x_3]$ \EndIf
    \State  \Return $[\sigma_{\max}^{-1}, \, \sigma_{\max} * x_1, \, \sigma_{\max} * x_2, \, \sigma_{\max} * x_3]$
\Else
    \State \Return $[\sigma_{\min}^{-1}, \, \sigma_{\min} * x_1, \, \sigma_{\min} * x_2, \, \sigma_{\min} * x_3]$
\EndIf
\EndProcedure
\end{algorithmic}
\end{algorithm}

\section{Normalizing vectors}
\label{secVectors}
Here we use algorithms Scale2D and Scale3D
to compute the length of two
and three dimensional vectors and normalize them accurately and efficiently.
This computation is performed by Algorithms \ref{algoNorm2} and
\ref{algoNorm3} in the end of this section.
Before the algorithms we present Lemma \ref{lem2D},
which describes their numerical properties. In summary,
Lemma \ref{lem2D} shows that, in an IEEE machine,
the scaling algorithms to normalize vectors incur in errors
which are only a small multiple of the unit roundoff, both in the
computed length of the vector and its direction.

Lemma \ref{lem2D} must be adapted for arithmetics
which do not follow the IEEE standard for floating
point arithmetic \cite{IEEE}. For instance, modern processors
allow for faster modes of execution called DAZ and FTZ,
which are incompatible with this standard.
When algorithms are executed in DAZ mode subnormal numbers are set
to zero on input to functions, and the input
$\wvec{x}$ in Lemma \ref{lem2D} must be considered to be what is left
after subnormals are set to zero,
but after this correction Lemma \ref{lem2D} does apply.
In FTZ mode subnormal numbers are set to zero on output
and the bounds on the angle in Lemma \ref{lem2D} would increase a bit, but would
still be of order $u$. The bound \pRef{b2a} would apply,
but in bound \pRef{b2b} we would need to replace $\wfpa/2$ by $\nu$.

\pbLemma{lem2D}
For $n = 2$ or $3$, suppose that Equations \pRef{condA}--\pRef{condF} are satisfied
and we execute Algorithms \ref{algoNorm2} and \ref{algoNorm3} with
input $\wvec{x} \in \wrn{n}$ and obtain $\hat{r} \in \wrone{}$ and $\hat{\wvec{x}} \in \wrn{n}$.
Let us define $r := \wvnorm{x}{}$ and $\phi \in [-\pi,\pi]$
as the angle between $\wvec{x}$ and $\hat{\wvec{x}}$.
If $\wvec{x} \neq 0$ is finite then $\hat{\wvec{x}}$ is also finite,
$\hat{r} \neq 0$,
\[
\wabs{\sin \phi} \leq 1.001 u
\hspace{1cm} \wrm{and} \hspace{1cm}
\wabs{\hat{\wvec{x}} - \wvec{x}/r} \leq \wlr{3.001 + n/2} u.
\]
If $\wlr{1 + \wlr{1 + n/2} u} r \leq \omega$ then $\hat{r}$ is finite and
\pbDef{b2a}
2 r \geq 3 \nu / 2 \, \Rightarrow  \,  \wabs{\hat{r} - r} \leq \wlr{1 + n / 2} r u
\peDef{b2a}
and
\pbDef{b2b}
2 r \leq 3 \nu / 2 \, \Rightarrow \,  \wabs{\hat{r} - r} \leq \wlr{1 + n / 2} r u + \wfpa/2.
\peDef{b2b}
\peLemma{lem2D}

\begin{algorithm}
\caption{Normalizing a two dimensional vector and computing its length}
\label{algoNorm2}
\begin{algorithmic}[1]
\Procedure{Normalize2D}{$x_1$, $x_2$}
\State \Comment Output: $[r, x_1 / r, \, x_2 / r]$ for $r = \sqrt{x_1^2 + x_2^2}$
\State $[\sigma, \tilde{x}, \tilde{x}_2] \gets \wflr{Scale2D}{x_1,x_2}$
\State
\If{ $\sigma = 0$ } \Return $[0,0,0]$ \EndIf
\State
\State $\tilde{r} \gets \sqrt{\tilde{x}_1 * \tilde{x}_1 + \tilde{x}_2 * \tilde{x}_2}$
\State $h \gets 1/\tilde{r}$
\State \Return $[\sigma * r, \, h * \tilde{x}_1 , \, h * \tilde{x}_2]$
\State
\EndProcedure
\end{algorithmic}
\end{algorithm}

\begin{algorithm}[H]
\caption{Normalizing a three dimensional vector and computing its length}
\label{algoNorm3}
\begin{algorithmic}[1]
\Procedure{Normalize3D}{$x_1$, $x_2$, $x_3$}
\State \Comment Output: $[r, x_1 / r, \, x_2 / r, x_3 / r]$ for $r = \sqrt{x_1^2 + x_2^2 + x_3^2}$.
\State $[\sigma, \tilde{x}, \tilde{x}_2, \tilde{x}_3] \gets \wflr{Scale3D}{x_1, \, x_2, \, x_3}$
\State
\If{ $\sigma = 0$ } \Return $[0,0,0,0]$ \EndIf
\State
\State $\tilde{r} \gets \sqrt{\tilde{x}_1 * \tilde{x}_1 + \tilde{x}_2 * \tilde{x}_2 + \tilde{x}_3 * \tilde{x}_3}$
\State $h \gets 1/\tilde{r}$
\State \Return $[\sigma * r, \, h * \tilde{x}_1, \, h * \tilde{x}_2, \, h * \tilde{x}_3]$
\State
\EndProcedure
\end{algorithmic}
\end{algorithm}

\section{Quaternions}
\label{secQuaternions}
In this section we present an scaling algorithm to normalize quaternions.
We are only concerned with the numerical aspects of computing with quaternions, and recommend
Altmann's book \cite{Altmann} for a broader discussion about them.
When the quaternion $\wvec{q} = \wlr{q_1,q_2,q_3,q_4}$ is different from zero
it is related to the rotation matrix
\[
\wvfc{R}{\wvec{q}} := \frac{1}{\wvnorm{q}{}^2}
\left(
\begin{array}{ccc}
q_1^2  + q_4^2 - q_2^2 - q_3^2 & 2 \wlr{q_1 q_2 - q_3 q_4}         & 2 \wlr{q_1 q_3 + q_2 q_4} \\
2 \wlr{q_1 q_2 + q_3 q_4}      & q_2^2 + q_4^2 - q_1^2 - q_3^2     & 2 \wlr{q_2 q_3 - q_1 q_4}   \\
2 \wlr{q_1 q_3 - q_2 q_4}      & 2 \wlr{q_2 q_3 + q_1 q_4}         & q_3^2 + q_4^2 - q_1^2 - q_2^2
\end{array}
\right),
\]
and when $\wvnorm{q}{} = 1$ this expression simplifies to
\[
\wvfc{R}{\wvec{q}} =
\left(
\begin{array}{ccc}
1 - 2 \wlr{q_2^2 + 2 q_3^2} & 2 \wlr{q_1 q_2 - q_3 q_4} & 2 \wlr{q_1 q_3 + q_2 q_4}  \\
2 \wlr{q_1 q_2 + q_3 q_4}   & 1 - 2 \wlr{q_1^2 + q_3^2} & 2 \wlr{q_2 q_3 - q_1 q_4}  \\
2 \wlr{q_1 q_3 - q_2 q_4}   & 2 \wlr{q_2 q_3 + q_1 q_4} & 1 - 2 \wlr{q_1^2 + 2 q_2^2}
\end{array}
\right).
\]
In the end of this section we present the scaling Algorithm \ref{algoNormQ},
which normalizes the quaternion $\wvec{q} = \wlr{x_1,x_2,x_3,x_4}$ so that
we can use the simplified expression for the rotation $\wvec{R}$ above.
The numerical properties of this algorithm are summarized by the following
lemma, which shows that the returned quaternion is normalized up to the
machine precision and the entries of the resulting matrix are also accurate,
provided, of course, that $\wvec{q}$ is finite and non zero.

\pbLemma{lemQ}
Suppose that Equations \pRef{condA}--\pRef{condF} are satisfied
and we execute Algorithm \ref{algoNormQ} with
input $\wvec{q}$ and obtain $\hat{r}$ and $\hat{\wvec{q}}$.
If $\wvec{q} \neq 0$ is finite then $\hat{\wvec{q}}$
is also finite, $\hat{r} \neq 0$ and
$\wnorm{\hat{\wvec{q}} - \bar{\wvec{q}}} \leq 5.001 u$,
for $\bar{\wvec{q}} := \wvec{q}/r$ and $r := \wvnorm{q}{}$. If $\wlr{1 + 3 u} r \leq \omega$ then $\hat{r}$ is finite and
\[
2 r \geq 3 \nu/2 \, \Rightarrow  \, \wabs{\hat{r} - r} \leq  3 r u
\]
and
\[
2 r \leq 3\nu/2 \, \Rightarrow  \, \wabs{\hat{r} - r} \leq  3 r u + \wfpa/2.
\]
Moreover, for $1 \leq i,j \leq 4$,
\[
\wabs{ \hat{q}_i \hat{q}_j - \bar{q}_i \bar{q}_j} \leq \wlr{1.001 + 8.001 \bar{q}_i \bar{q}_j} u.
\pLink{lemQ}
\]
\peLemmaX{lemQ}

\begin{algorithm}
\caption{Normalizing a quaternion and computing its length}
\label{algoNormQ}
\begin{algorithmic}[1]
\Procedure{Normalize4D}{$x_1$, $x_2$, $x_3$, $x_4$}
\State \Comment Output: $[r, x_1/r, \, x_2/ r, x_3/ r, x_4/r]$ for $r = \sqrt{x_1^2 + x_2^2 + x_3^2 + x_4^2}$.
\State $\sigma \gets 0$ \Comment Just a formality for the proofs
\State $m \gets \wabs{x_1}$
\If{ $m < \wabs{x_2}$ }
    \State $m \gets \wabs{x_2}$
    \If{ $m < \wabs{x_3}$  }
        \State $m \gets \wabs{x_3}$
    \EndIf
    \If{ $m < \wabs{x_4}$  }
        \State $m \gets \wabs{x_4}$
    \EndIf
\Else
    \If{ $m < \wabs{x_3}$  }
        \State $m \gets \wabs{x_3}$
        \If{ $m < \wabs{x_4}$  }
            \State $m \gets \wabs{x_4}$
        \EndIf
    \Else
        \If{ $m \geq \wabs{x_4}$ }
           \If{ $m = 0$ } \Return $[0,0,1,0]$ \EndIf
           \State $m \gets\wabs{x_4}$
        \EndIf
    \EndIf
\EndIf
\State
\If{ $m \geq \tau_{\min}$ }
    \If{ $m \leq \tau_{\max}$ }
        \State $\sigma \gets 1$, $\ \ \ \tilde{x}_1 \gets x_1$, $\ \ \ \tilde{x}_2 \gets x_2$, $\ \ \ \tilde{x}_3 \gets x_3$, $\ \ \ \tilde{x}_4 \gets x_4$,
    \Else
        \State $\sigma \gets \sigma_{\max}^{-1}$, $\ \ \  \tilde{x}_1 \gets \sigma_{\max} * x_1$, $\ \ \  \tilde{x}_2 \gets \sigma_{\max} * x_2$,
        \State $\tilde{x}_3 \gets \sigma_{\max} * x_3$, $\ \ \  \tilde{x}_4 \gets \sigma_{\max} * x_4$
    \EndIf
\Else
\State $\sigma \gets \sigma_{\min}^{-1}$, $\ \ \  \tilde{x}_1 \gets \sigma_{\min} * x_1$, $\ \ \  \tilde{x}_2 \gets \sigma_{\min} * x_2$,
\State $\tilde{x}_3 \gets \sigma_{\min} * x_3$, $\ \ \  \tilde{x}_4 \gets \sigma_{\min} * x_4$
\EndIf
\State
\State $\tilde{r} \gets \sqrt{\tilde{x}_1 * \tilde{x}_1 + \tilde{x}_2 * \tilde{x}_2 + \tilde{x}_3 * \tilde{x}_3 + \tilde{x}_4 * \tilde{x}_4}$
\State $h \gets 1/\tilde{r}$
\State \Return $[\sigma * r, \, h * \tilde{x}_1, \, h * \tilde{x}_2, \, h * \tilde{x}_3, \, h * \tilde{x}_4]$
\EndProcedure
\end{algorithmic}
\end{algorithm}

\appendix
\section{Experiments}
\label{secExperiments}
The experimental results are summarized in tables \ref{tableTimes} and
\ref{tablePlain},
and after these tables we explain how the experiments were performed.
Table \ref{tableTimes} reports the ratio of the time taken by quotient
algorithm and the time taken by scaling algorithm to execute the task
mentioned in the first column. Each entry in Table \ref{tableTimes}
is the average of a  fairly large number of experiments,
and the number after the $\pm$ sign is the
standard deviation.

\begin{table}
\caption{$\frac{\wrm{Time \ taken\  by \ a  \ quotient \ algorithm}}{\wrm{Time \ taken \ by \ the \ corresponding  \ scaling \ algorithm}}$}
\centering
\begin{tabular}{cc|ccc}
             &              &  \multicolumn{3}{c}{Compiler/Operating system} \\[0.02cm]
             &              & GCC     &  Intel  C++  & Visual Studio   \\
Task         &  Precision   & Linux   &  Linux       & Windows         \\
\hline
Normalizing &   single    &  $1.69 \pm 0.07$  & $1.72 \pm 0.03$ & $2.08 \pm 0.12$ \\
a 2D vector &  double     &  $2.46 \pm 0.05$  & $1.92 \pm 0.03$ & $2.56 \pm 0.11$ \\
\hline
Normalizing  & single     &  $1.70 \pm 0.04$  & $1.95 \pm 0.03$ & $2.31 \pm 0.16$  \\
a 3D vector  & double     &  $2.53 \pm 0.02$  & $2.59 \pm 0.08$ & $2.99 \pm 0.19$ \\
\hline
Normalizing   & single     &  $1.77 \pm 0.07$  & $1.83 \pm 0.02$ & $1.96 \pm 0.10$   \\
a Quaternion  & double     &  $2.54 \pm 0.12$  & $2.60 \pm 0.03$ & $2.80 \pm 0.14$ \\
\end{tabular}
\label{tableTimes}
\end{table}

\begin{table}[h]
\caption{$\frac{\wrm{Time \ taken\  by \ a \ scaling \ algorithm}}{\wrm{Time \ taken \ by \ the \ naive \ algorithm}}$}
\centering
\begin{tabular}{cc|ccc}
             &              &  \multicolumn{3}{c}{Compiler/Operating system} \\[0.02cm]
             &              & GCC     &  Intel  C++  & Visual Studio   \\
Task         &  Precision   & Linux   &  Linux       & Windows         \\
\hline
Normalizing &   single   &  $1.15 \pm 0.05$  & $1.04 \pm 0.02$ & $1.08 \pm 0.07$ \\
a 2D vector &  double    &  $1.17 \pm 0.02$  & $1.01 \pm 0.02$ & $1.00 \pm 0.05$ \\
\hline
Normalizing  & single    &  $1.38 \pm 0.05$ & $1.19 \pm 0.02$ & $1.22 \pm 0.09$  \\
a 3D vector  & double    &  $1.36 \pm 0.01$ & $1.08 \pm 0.53$ & $1.39 \pm 0.11$ \\
\hline
Normalizing   & single   &  $1.38\pm0.06$  & $1.89 \pm 0.13$  & $1.46 \pm 0.10$  \\
a Quaternion  & double   &  $1.42\pm0.08$  & $1.38 \pm 0.02$  & $1.66 \pm 0.09$ \\
\end{tabular}
\label{tablePlain}
\end{table}

The experiments used an Intel Core i7 2700K cpu
with Ubuntu 14.04 LTS and an
Intel Core i7 950 cpu with Windows 10 Pro, version 1511.
The code for the experiments was written in
C++11 and was compiled with GCC 4.9.3 and Intel C++ 16.0.1
in Linux, with the -O3 optimizing flag (but not the flag -ffast-math.)
In windows we used Visual Studio Enterprise 2015, version 14.0.25123.00,
update 2, with the standard optimization flags for release builds.
We tried to write simple code, with no effort to tune any algorithm
to any machine. This code is available upon request to the author. Each entry in
Tables \ref{tableTimes} and \ref{tablePlain} is the average over 500 experiments, and in each
experiment we executed the algorithm $10^{6}$ times, in loops with a bit
of extra code so that the optimizer would not remove the function
calls that we wanted to time. We timed the loops without the function
calls and subtracted the time for the empty loops from the ones
we wanted to time. We measured the time with the functions
\verb clock_gettime  in Linux and \verb GetProcessTimes in Windows,
and took into account only the time taken by the process.
\section{Proofs}
\label{secProofs}
In this section we prove Lemmas \ref{lem2D} and \ref{lemQ}.
Our proofs are based on \cite{RealNumbers}, but
one could obtain similar results using the theory in \cite{Higham}.
However, we note that \cite{RealNumbers} takes underflow into account
whereas the standard model of floating point arithmetic in \cite{Higham} does not,
and dealing with underflow and overflow is the motivation
for this article. As \cite{Higham}, we use $\wfl{\wrm{expr}}$ to denote the
computed value for the expression $\wrm{expr}$.

Our proofs use the following well known facts about IEEE floating point
arithmetic, in which $x$ is a finite floating point number, $z \in \wrone{}$
and $\beta$ is a power of the base:
\begin{eqnarray}
\pClaimLabel{rndA}
 \nu \leq \wabs{\beta x} \leq \omega    & \Rightarrow & \wfl{\beta x} = \beta x, \\
\pClaimLabel{rndB}
 \wabs{\beta x} \leq 2 \nu    & \Rightarrow & \wfl{\beta x} = \beta x + \delta \wfpa \ \wrm{with} \ \ \wabs{\delta} \leq 1/2,  \\
 \pClaimLabel{rndC}
 \wfl{\beta x}  & = & \beta x  + \delta \wfpa \ \ \wrm{with} \ \ \wabs{\delta} \leq 1/2, \\
 \pClaimLabel{rndD}
 \nu \leq \wabs{z} \leq \omega    & \Rightarrow & \wabs{\wfl{z} - z} \leq \frac{u}{1 + u} \wabs{z}, \\
  \pClaimLabel{rndE}
 \wabs{z} \leq 2 \nu  & \Rightarrow  & \wfl{z} = z + \delta \wfpa \ \ \wrm{with}  \ \wabs{\delta} \leq 1/2, \\
 \pClaimLabel{rndF}
    \wfl{z}  & = & \gamma z  + \delta \wfpa \ \ \wrm{with} \ \
    \wabs{\gamma - 1} \leq \frac{u}{1 + u} \ \ \wrm{and} \ \ \wabs{\delta} \leq 1/2.
\end{eqnarray}
We now state and proof some auxiliary results and after that
we present the proofs of Lemmas \ref{lem2D} and \ref{lemQ}.
%
%
\pbLemma{lemSquares}
For $n = 2,3,4$, let $\sigma$ and $\tilde{x}_k$, $k = 1, \dots, n$
be the numbers in Algorithms \ref{algoNorm2}, \ref{algoNorm3}
and \ref{algoNormQ} with input $x_k$, $k = 1,\dots,n$
and define $\check{x}_k := \sigma^{-1} x_k$. If the $x_k$ are finite
and $\sigma \neq 0$ then $\wfl{\tilde{x}_k^2} = \wfl{\check{x}_k^2}$.
\peLemma{lemSquares}
%
%

%
%

\pbProofB{Lemma}{lemSquares}
When $\sigma \neq 0$ we have that
$\tilde{x}_k = \wfl{\check{x}_k}$. Since $\sigma$ is a power of $2$,
if $\wabs{\check{x}_k} \geq \nu$ then
$\tilde{x}_k = \wfl{\check{x}} = \check{x}_k$ because $\check{x}$ is a
floating point number in this case. Therefore,
$\wfl{\tilde{x}_k^2} = \wfl{\check{x}^2}$ when $\wabs{\check{x}} \geq \nu$.
If $\wabs{\check{x}} < \nu$ then Equation \pRef{condB} and the monotonicity of rounding
lead to
$\wabs{\tilde{x}_k} = \wabs{\wfl{\check{x}}} \leq \nu$ and
$0 \leq \wfl{\tilde{x}_k^2}, \wfl{\check{x}^2} \leq \wfl{\nu^2} = 0$.
\peProof{Lemma}{lemSquares}\\

%
%

%
%

\pbLemma{lemXHat}
For $n = 2$, $3$ or $4$, assume that $\wvec{x} \in \wrn{n}$ and
that Equations \pRef{condA}--\pRef{condF} hold.
If we execute Algorithms \ref{algoNorm2}, \ref{algoNorm3} or \ref{algoNormQ} with
a finite input $\wvec{x} \neq 0$ then the returned $\hat{r}$ and $\hat{\wvec{x}}$  are
such that $\hat{r} \neq 0$ and, for $r := \wvnorm{x}{}$ and $\bar{\wvec{x}} := \wvec{x}/r$,
\[
\hat{\wvec{x}} = \lambda \wlr{\bar{\wvec{x}} + \wvec{w} + \wvec{z}}
\hspace{1.5cm} \wrm{with} \hspace{1.5cm}
\wabs{\lambda - 1} \leq \wlr{2.001 + n/ 2} u,
\]
\[
\wabs{w_k} \leq u \wabs{\bar{x}_k}/\wlr{1 + u}
\hspace{1.5cm} \wrm{and} \hspace{1.5cm}
\wvnorm{z}{} \leq 2 \sqrt{\wfpa} u.
\]
If $\wlr{1 + \wlr{1 + n/2} u} r \leq \omega$ then $\hat{r}$ is finite and
\pbDef{bndrnA}
2 r \geq 3 \nu \ \Rightarrow \ \wabs{\hat{r} - r} \leq \wlr{1 + n/2}  r u
\peDef{bndrnA}
and
\pbDef{bndrnB}
2 r \leq 3 \nu \ \Rightarrow \ \wabs{\hat{r} - r} \leq \wlr{1 + n/2} r u + \wfpa/2.
\peDef{bndrnB}
Finally,
\[
\wvnorm{\hat{x} - \bar{x}}{} \leq \wlr{3.001 + n/2} u.
\pLink{lemXHat}
\]
\peLemmaX{lemXHat}

%
%

%
%

\pbProof{Lemma}{lemXHat}
Let $\ell$ be an index such that $\wabs{x_\ell} = \max \wabs{x_k}$.
In Algorithms \ref{algoScale2}, \ref{algoScale3} and \ref{algoNormQ},
$\sigma \neq 0$ because $x_{\ell} \neq 0$. Let
$\tilde{x}_k$ be the numbers in these algorithms and define
\[
s := r^2,
\hspace{1.0cm}
\tilde{r} := \wnorm{\tilde{x}}{},
\hspace{1.0cm}
\tilde{s} := \tilde{r}^2,
\hspace{1.0cm}
\wrm{and} \hspace{1.0cm}
\check{\wvec{x}} := \sigma^{-1} \wvec{x}.
\]
Equations \pRef{condA}--\pRef{condF} imply
that $\check{x}_{\ell} \in [\tau_{\min},\tau_{\max}]$,
 $\wabs{\check{x}_{\ell}} \in [\nu,\omega]$,
$\tilde{x}_{\ell} = \check{x}_{\ell}$ and
$\tilde{x}_k^2 \leq \tilde{x}_{\ell}^2 \leq \omega/8$.
It follows that $\nu \leq \wfl{\tilde{r}} < \omega$ and
Corollary 2 in \cite{RealNumbers} implies that
\pbDef{frt}
\wfl{\tilde{r}} \geq \wfl{\sqrt{\tilde{x}_{\ell}^2}} = \wabs{\tilde{x}_{\ell}} = \sigma^{-1} \wabs{x_{\ell}} \geq \tau_{\min}.
\peDef{frt}
Lemma \ref{lemSquares} shows that
\[
\wfl{\tilde{s}} = \wfl{ \sigma^{-2} s} \geq \tau_{\min}^2 \geq \wfpa / u^2,
\]
and the argument used in the proof of Corollary 11 in \cite{RealNumbers} and a bit of Algebra lead to
\[
\wfl{\tilde{s}} = \gamma_s \sigma^{-2} s
\hspace{1cm} \wrm{with} \hspace{1cm}
\wabs{\gamma_s - 1} \leq n u
\]
(Note that $n+1$ in \cite{RealNumbers} correspond to $n$ here.)
As a result,
\[
\wfl{\tilde{r}} = \wfl{ \sqrt{ \wfl{\tilde{s}}}} =
\gamma_r \sqrt{\gamma_s} \sigma^{-1} r
\hspace{1cm} \wrm{with} \hspace{1cm} \wabs{\gamma_r - 1} \leq \frac{u}{1 + u}.
\]
The convexity/concavity arguments in \cite{RealNumbers} lead to
\[
\gamma := \gamma_r \sqrt{\gamma_s}
\hspace{1cm} \wrm{satisfies} \hspace{1cm}
\wabs{\gamma - 1} \leq \wlr{1 + n/2} u,
\]
and Equation \pRef{condD} leads to
\pbDef{tr}
\wfl{\tilde{r}} = \wfl{ \sqrt{ \wfl{\tilde{s}}}} = \gamma \sigma^{-1} r \leq
2 \gamma \sigma^{-1} \wabs{x_{\ell}} \leq 3 \tau_{\max} \leq 1/\nu.
\peDef{tr}
If $3 \nu / 2 \leq r \leq \omega / \gamma$ then Equation \pRef{tr} leads to
\[
\omega \geq \sigma \wabs{\wfl{\tilde{r}}} \geq 3 \gamma \nu/2 > \nu,
\]
Equations \pRef{rndA} and \pRef{frt} lead to
\[
0 < \wabs{x_{\ell}} \leq \hat{r} = \wfl{\sigma \wfl{\tilde{r}}} = \sigma \wfl{\tilde{r}} \leq \omega,
\]
and Equation \pRef{tr} shows that
$\wabs{\hat{r} - r} \leq \wabs{\gamma - 1} r$,
and Equation \pRef{bndrnA} holds. The bound \pRef{bndrnB}
follows from the argument above and Equation \pRef{rndE}.

In algorithms \ref{algoNorm2}, \ref{algoNorm3} and \ref{algoNormQ}, $\tilde{x}_k = \wfl{\sigma^{-1} x_k}$ and
Equation \pRef{rndC} implies that
\pbDef{xtil}
\tilde{x}_k = \sigma^{-1} x_k + \rho_k \wfpa \ \ \wrm{with} \ \ \wabs{\rho_k} \leq 1/2.
\peDef{xtil}
We also have $h = \wfl{1/\wfl{\tilde{r}}}$ and Equations
\pRef{rndD} and \pRef{tr} lead to
\[
h = \frac{\sigma \lambda}{r}
\hspace{0.5cm} \wrm{with} \hspace{0.5cm}
\lambda := \gamma_h / \gamma = \frac{\gamma_h}{\gamma_r \sqrt{\gamma_s}}
\hspace{0.5cm} \wrm{and} \hspace{0.5cm}
\wabs{\gamma_h - 1} \leq \frac{u}{1 + u}.
\]
The usual upper bound on $\lambda = \gamma_h / \gamma$
is not a concave function of $u$, but since $u < 10^{-6}$ we can prove that
\pbDef{lamb}
\wabs{\lambda - 1} \leq \wlr{2.0001 + n/2} u,
\peDef{lamb}
and this bound is a bit better than what is claimed by Lemma \ref{lemXHat}.
Since $\hat{x}_k = \wfl{h \tilde{x}_k}$ in all algorithms,
Equations \pRef{rndF} and \pRef{xtil} yield
\[
\hat{x}_k = \lambda \gamma_k  \wlr{\bar{x}_k + \rho_k \frac{\sigma \wfpa}{r}} + \eta_k \wfpa
\hspace{0.25cm} \wrm{with} \hspace{0.25cm}
 \wabs{\gamma_k - 1} \leq \frac{u}{1+u}
\hspace{0.25cm} \wrm{and} \hspace{0.25cm}
\wabs{\eta_k} \leq 1/2,
\]
and $\hat{x}_k = \lambda \wlr{\bar{x}_k + w_k + z_k}$ with
\pbDef{wkzk}
w_k := \wlr{\lambda_k - 1} \bar{x}_k
\hspace{0.5cm} \wrm{and} \hspace{0.5cm}
z_k := \gamma_k \rho_k \sigma \wfpa/ r + \eta_k \wfpa/\lambda.
\peDef{wkzk}
Since $r \geq \wabs{x_\ell}$ and $\tau_{\min} \geq \sqrt{\wfpa}/u$ we have that
\[
\wfpa \sigma/r \leq \wfpa/\wlr{\sigma^{-1} \wabs{x_\ell}} \leq \wfpa/\tau_{\min}  \leq \sqrt{\wfpa} u ,
\]
and the bounds on $\gamma_k$, $\rho_k$, $\eta_k$ and $\lambda$ above
and Equation \pRef{condA} imply
that $\wvnorm{z}{} \leq 2 \sqrt{\wfpa} u$.
Finally, since $\wvnorm{\bar{x}}{} = 1$, the bounds above lead to
\[
\wvnorm{\hat{x} - \bar{x}}{}
\leq \wabs{\lambda - 1}\, \wvnorm{\bar{x}}{} + \lambda \, \wvnorm{w}{} + \lambda \, \wvnorm{z}{}
\]
\[
\leq \wlr{2.0001 + n/2} u + \wlr{1 +  \wlr{2.0001 + n/2} u} \frac{u}{1 + u} + 2 \lambda \sqrt{\wfpa} u
\leq \wlr{3.001 + n/2} u,
\]
and this completes the proof of Lemma \ref{lemXHat}.
\peProof{Lemma}{lemXHat} \\

%
%

%
%

\pbProofB{Lemma}{lem2D}
The only part of Lemma \ref{lem2D} which does not follow directly from
Lemma \ref{lemXHat} is the bound on $\phi$, which we now verify.
Equation \pRef{wkzk} and the bounds on $\lambda_k$ and $z_k$ in the
proof of Lemma \ref{lemXHat} lead to
\[
\wvnorm{\hat{x}{}}{} \geq \lambda \wlr{1 - \frac{u}{1 + u} - 2 \sqrt{\wfpa} u }
= \frac{\lambda}{1 + u} \wlr{1 - 2 \wlr{1 + u} \sqrt{\wfpa} u}
\]
and the usual formula for the cross product leads to
\[
\wabs{\sin \phi} =
\frac{\wnorm{\wvec{\bar{x}} \times \hat{\wvec{x}}}}{\wvnorm{\bar{x}}{} \wvnorm{\hat{x}}{}}
\leq
\wlr{1 + u} \frac{\wnorm{\bar{\wvec{x}} \times \wlr{\wvec{w} + \wvec{z}}}}{1 - 2 \wlr{1 + u} \sqrt{\wfpa} u }
\leq
\wlr{1 + u} \frac{\wvnorm{w}{} + \wvnorm{z}{}}{1 - 2 \wlr{1 + u} \sqrt{\wfpa} u }
\]
\[
\leq \frac{u + 2 \wlr{1 + u} \sqrt{\wfpa} u}{1 - 2 \wlr{1 + u} \sqrt{\wfpa} u }
\leq 1.001 u,
\]
and we are done.
\peProof{Lemma}{lem2D}\\

%
%

%
%

\pbProofB{Lemma}{lemQ}
The only part of Lemma \ref{lemQ} which does not follow directly from
Lemma \ref{lemXHat} with $n = 4$ is the bound on $q_i q_j$, which we now verify.
Equation \pRef{wkzk} and the bounds on $\lambda_k$ and $z_k$ in the
proof of Lemma \ref{lemXHat} lead to
\[
\hat{q}_i \hat{q}_j - \bar{q}_i \bar{q}_j = \wlr{\lambda^2 - 1} \bar{q}_i \bar{q}_j + \lambda^2 \delta
\]
for
\[
\delta := \bar{q}_i \wlr{w_j + z_j} + \bar{q}_j \wlr{w_i + z_i} + \wlr{w_i + z_i} \wlr{w_j + z_j}.
\]
Equation \pRef{lamb} with $n = 4$ and the fact that $\wvnorm{q}{} = 1$ lead to
\pbDef{qf1}
\wabs{\wlr{\lambda^2 - 1} \bar{q}_i \bar{q}_j} \leq
\wabs{\lambda - 1} \wabs{\lambda + 1} \leq
4.0001 u \wlr{2 + 4.0001 u} \leq 8.001 u
\peDef{qf1}
and
\[
\wabs{\delta} \leq \wvnorm{\bar{q}}{} \wvnorm{w + z}{} + \wlr{\wvnorm{w}{} + \wvnorm{z}{}}^2
\leq \wvnorm{w}{} + \wvnorm{z}{} + \wlr{\wvnorm{w}{} + \wvnorm{z}{}}^2
\]
\[
\leq \frac{u}{1 + u} + 2 \sqrt{\wfpa} u + \wlr{\frac{1}{1 + u} + 2 \sqrt{\wfpa} }^2 u^2 \leq u.
\]
Therefore
\pbDef{qf2}
\lambda^2 \wabs{\delta} \leq 1.001 u
\peDef{qf2}
and the bound on $\wabs{\hat{q}_i \hat{q}_j - \bar{q}_i \bar{q}_j}$ follows from
Equations \pRef{qf1} and \pRef{qf2}.
\peProof{Lemma}{lemQ}\\

%
%

\end{document}